\documentclass[amsmath,amssymb,prd,twocolumn,nofootinbib]{revtex4}

\usepackage{graphicx}
\usepackage{color}
\usepackage{subfigure}

\usepackage{citesort}
\include{aas_journals}

\newcommand{\be}{\begin{equation}}
\newcommand{\ee}{\end{equation}}
\newcommand{\bea}{\begin{eqnarray}}
\newcommand{\eea}{\end{eqnarray}}
\newcommand{\bean}{\begin{eqnarray*}}
\newcommand{\eean}{\end{eqnarray*}}

\begin{document}
\title{Constraints on the Brans-Dicke gravity theory with the 
{\it Planck} data}
\author{Yi-Chao Li$^{1,2}$}
\email{ycli@bao.ac.cn}
\author{Feng-Quan Wu$^1$ }
\email{wufq@bao.ac.cn}
\author{Xuelei Chen$^{1,3}$}
\email{xuelei@cosmology.bao.ac.cn}
\affiliation{$^1$National Astronomical Observatories, Chinese
Academy of Sciences, \\
20A Datun Road, Chaoyang District, Beijing 100012, China\\
$^2$ University of Chinese Academy of Sciences, 
Beijing 100049, China \\
$^3$ Center of High Energy Physics, Peking University, Beijing 100871, China
}

\date{\today}

\begin{abstract}

Based on the new cosmic microwave background (CMB) temperature data 
from the {\it Planck} satellite, the 9 year polarization data from the 
Wilkinson Microwave Anisotropy Probe (WMAP), 
and the baryon acoustic oscillation (BAO) distance ratio data 
from the Sloan Digital Sky Survey (SDSS) and 6 Degree field (6dF) surveys, 
we place a new constraint on the Brans-Dicke theory. We adopt a parametrization 
$\zeta=\ln(1+\frac{1}{\omega})$, where the general relativity (GR) 
limit corresponds to 
$\zeta = 0$. We find no evidence of deviation from general relativity. 
At 95\% probability, $-0.00246 < \zeta < 0.00567$, correspondingly, the region  
$-407.0 < \omega <175.87$ is excluded. If we restrict ourselves to the $\zeta>0$ 
(i.e. $\omega >0$) case, then the 95\% probability interval is $\zeta<0.00549$, 
corresponding to $\omega> 181.65$. 
We can also translate this result to a constraint
on the variation of gravitational constant, and find the variation rate today as
$\dot{G}=-1.42^{+2.48}_{-2.27}~\times 10^{-13} $yr$^{-1} $ 
( $1\sigma$ error bar), 
the integrated change since the epoch of 
recombination is $\delta G/G = 0.0104^{+0.0186}_{-0.0067} $  ($1\sigma$ error bar). 
These limits on the variation of gravitational constant 
are comparable with the precision of solar system experiments.

\end{abstract}


\maketitle


\section{Introduction}
The Jordan-Fierz-Brans-Dicke theory 
\cite{Jordan:1949nature,Jordan:1959eg,Fierz:1956,Brans:1961sx,Dicke:1961gz}
(hereafter the Brans-Dicke theory for simplicity)
is the simplest extended theory of gravity. 
In addition to the metric tensor, there is a scalar field $\phi$ 
in this theory, 
i.e. the Brans-Dicke field, which gives the effective gravitational 
constant\cite{Bergmann:1968ve,Nordtvedt:1970uv,Wagoner:1970vr,Bekenstein:1977rb,
Bekenstein:1978zz}. The action in the Jordan frame is
\begin{equation}
\label{action} 
{\mathcal S} = \frac{1}{16\pi} \int d^4 x \sqrt{-g}
\left[-\phi R + \frac{w}{\phi} g^{\mu\nu} \nabla_{\mu} \phi \nabla_{\nu} \phi \right]
+ {\mathcal S}^{(m)} \ ,
\end{equation}
where ${\mathcal S}^{(m)}$ is the action for the matter field.
Here $\phi$ is the Brans-Dicke field and $\omega$ is the 
Brans-Dicke parameter. 
In the limit of $\omega \to \infty$ the Brans-Dicke theory is reduced to 
Einstein's general relativity theory.

Solar system experiments have already put very stringent 
constraints on the Brans-Dicke model\cite{Will:1981tegp.book, Will:2006wv}. 
For example, the tracking data obtained from the Cassini mission 
gives $\omega > 40000$ at the $2\sigma$ level \cite{Bertotti:2003by}. Nevertheless,
it is still interesting to test the theory on cosmological scales, especially 
because the Brans-Dicke theory may be regarded as an approximation for 
a number of scalar-tensor theories of gravity which have more significant effects
on larger scales. The cosmic microwave background (CMB) anisotropy can be calculated
for a given theory, and the Brans-Dicke model may be tested with high precision
\cite{Chen:1999qh}.

A number of limits on the $\omega$ parameter have been derived since 
WMAP released its data on CMB anisotropy.
Acquaviva et al. report that $\omega > 80$ at $99\%$ level by combining the 
WMAP 1st year data and some ground or balloon based 
experiments and the large scale structure data\cite{Acquaviva:2004ti}.  
Wu et al. 
\cite{wu:2010brans.theory,wu:2010brans.constraint} excluded the region of  
$-120 < \omega < 97.8$ by using the
WMAP 5 year data and the Sloan Digital Sky Survey (SDSS) 
Luminous Red Galaxy (LRG) 
data.  Considering only the possibility of $\omega >0$, 
Avilez and Skordis\cite{Avilez:2013tm} derived a limit 
of  $\omega > 288$ at $95\%$ confidence level. by combining the
WMAP 7 year data and the data from  
the South Pole Telescope (SPT) and other small scale CMB experiments. 
When comparing these different results, one should note
that the limits obtained depend very much 
on the parameterization and prior used, 
see the next section for discussion.

The precision of cosmological observations are being improved steadily. The 
WMAP group have published the data of 9 years of 
observation \cite{2012arXiv1212.5225B,2012arXiv1212.5226H}, and recently,
the Planck collaboration published their observational 
results \cite{2013arXiv1303.5062P}. In addition to the CMB observations, 
there are also much progress in redshift surveys of galaxies.
Recent surveys such as the SDSS-III Baryon Oscillation Spectroscopic 
Survey (BOSS) \footnote{http://www.sdss3.org/surveys/boss.php} and 
6dF \footnote{http://www.aao.gov.au/6dFGS/} have measured the 
power spectrum of the large scale structure at different redshifts, and obtained
cosmic distances from the baryon acoustic oscillation features. 

In this paper, we update the constraint on the Brans-Dicke 
model by using the new CMB data \footnote{http://lambda.gsfc.nasa.gov/}, 
including the Planck temperature anisotropy \cite{collaboration:2013vc} and the 
the WMAP9 CMB polarization data\cite{2012arXiv1212.5225B}. Following the Planck 
collaboration \cite{2013arXiv1303.5076P}, in addition to the CMB data, 
we also use the BAO data from 
the SDSS \cite{Percival:2010hx,Padmanabhan:2012ft,Anderson:2012uh} 
and 6dF \cite{Beutler:2011ea} galaxy redshift surveys. 

\section{Methods}

For convenience, we introduce a dimensionless 
field $\varphi = G\phi$ where $G$ is the Newtonian gravitational constant, 
then the Einstein equations are 
\begin{eqnarray}
\label{generalized Einstein equation}
G_{\mu\nu} &=& \frac{8 \pi G}{\varphi}T^{(m)}_{\mu\nu}
+ \frac{\omega}{{\varphi}^2} (\nabla_{\mu}\varphi\nabla_{\nu}\varphi
- \frac{1}{2}g_{\mu\nu}\nabla_{\lambda}\varphi\nabla^{\lambda}\varphi)
\nonumber \\ &&
+ \frac{1}{\varphi}(\nabla_{\mu}\nabla_{\nu}\varphi
- g_{\mu\nu}\nabla_{\lambda}\nabla^{\lambda}\varphi) \ ,
\end{eqnarray}
where $T^{(m)}_{\mu\nu}$ is the stress tensor for matter. The 
equation of motion for $\varphi$ is 
\begin{equation}
\nabla_a\nabla^a\varphi = \frac{\kappa}{2\omega+3}T_{\nu}^{(m)\mu} \ .
\end{equation}
For $G$ to be consistent with the Cavendish experiment, 
The value of $\varphi$ at present day should be
\begin{equation}
\varphi _0 = \frac{2\omega+4}{2\omega+3} \ .
\end{equation}

We follow the calculation method 
described in \cite{wu:2010brans.theory}, in which we developed the covariant and
gauge-invariant formalism of cosmological perturbation theory 
in the case of Brans-Dicke gravity, and apply the method 
to calculate the angular power spectra of CMB temperature and polarizations,
as well as the power spectrum of  large scale structure (LSS).

Given a cosmological model, the angular power spectra of CMB temperature and 
polarization and the matter power spectrum 
can be calculated, for example with the publicly available code
{\tt CAMB} \cite{CAMB}. 
In order to constrain the cosmological parameters with the observational 
data, we use the publicly available CosmoMC code \cite{Lewis:2002ah}, which 
uses the Markov Chain Monte Carlo (MCMC) method to explore the parameter space, 
with a modified {\tt CAMB} code developed by Wu et al. in Ref \cite{wu:2010brans.constraint}. 
We use the latest CMB data published 
by the Planck team \cite{collaboration:2013vc}. 
According to our previous analysis \cite{wu:2010brans.theory}, 
the small scale (high-$l$) anisotropy is affected more by the Brans-Dicke 
gravity, so the more precise measurements of {\it Planck} on small scales (up to 
$l \sim 2500 $) should help greatly.  For the low-$l$s, 
we also include the TE and BB power spectrum estimated from the  
polarization map of WMAP9, though the latter does not provide much distinguishing power
at present. 

We also combine the BAO data from large scale structure surveys, including the
SDSS DR7 \cite{Percival:2010hx, Padmanabhan:2012ft}, 
BOSS DR9 \cite{Anderson:2012uh} and 6dF \cite{Beutler:2011ea}. The BAO surveys measure 
the distance ratio
\begin{equation}
d_z = \frac{r_s(z_{drag})}{D_V(z)} \ ,
\end{equation}
where $r_s(z_{drag})$ is the comoving sound horizon when baryons became dynamically
decoupled from photons (the baryon drag epoch) and $D_V(z)$ is the combination of
angular-diameter distance, $D_A(z)$, and the Hubble parameter, $H(z)$
\begin{equation}
  D_V(z) = [(1+z)^2D_A^2(z)\frac{cz}{H(z)}]^{1/3} \ .
\end{equation}
We follow the choice of BAO data set ``SDSS DR7 + BOSS DR9 + 6dF" in the Planck 
analysis \cite{collaboration:2013vc}. This includes two of the most accurate BAO 
measurements, and minimizes the correlations between the galaxy surveys, as the two
surveys have widely separated effective redshifts.

The derived limits depend on which parameterization is selected and 
how the priors are set. As the
experiments so far all favors the GR case and the Brans-Dicke parameter is 
stringently constrained, it is more convenient to take the GR as the 
null case, and have a parametrization in which the tested parameter vanishes for 
GR. In practice a flat prior on a finite range is usually assigned to the parameter.  
For example, Ref.\cite{Acquaviva:2004ti} and Ref.\cite{Avilez:2013tm} 
considered flat prior on $\ln[1/\omega]$ (though the exact 
parameter they used differs slightly). A limitation of this choice is that 
it could not treat the $\omega<0$ case. Ref.\cite{Avilez:2013tm} 
argued that if $\omega < -3/2$ the Brans-Dicke field would be a ghost field, and 
they will therefore consider only positive $\omega$. However, we would rather 
err on the conservative side, and use a more general form of parameterization
which allows negative $\omega$. Indeed, at present there are many phantom dark 
energy \cite{2003PhRvL..91g1301C} models in which the field 
are also ghost-like. In Ref.\cite{wu:2010brans.constraint} 
we used 
\begin{equation}
\zeta = \ln(1+\frac{1}{\omega}) \ ,
\end{equation}
in the present paper we will also adopt this parameterization.
This parameter has the nice property that as $\zeta \to 0$, the 
Brans-Dicke theory reduces to Einstein 
gravity, and it is easy to obtain limits on both the negative and 
positive value of $\omega$. We choose the same initial range $[-0.014, 0.039]$ 
as Wu et al. in \cite{wu:2010brans.constraint}, 
which is convenient for computation, while at the same time the final 
constraint is not very sensitive to this range, since at the edge of the 
prior range the likelihood is very small. In fact, if one wishes to consider only 
positive values of $\omega$, we can also do that by simply restricting 
the range of the prior to $0<\zeta < 0.039$.

As pointed out by Ref.\cite{Avilez:2013tm}, comparing 
with their prior, our flat prior on $\zeta$ penalizes
large $\omega$, and hence for the same data set a ``weaker''
limit on $\omega$ would be obtained for our choice. We do not see a good 
theoretical reason to favor one prior on $\omega$ over the other, 
but our choice is again in agreement with our general 
philosophy of being conservative on 
constraining models. We remind the reader to notice the 
effect of the prior when comparing results obtained in different papers.

We also obtain limits on the following basic or derived cosmological parameters:
$\Omega_{\Lambda}$, $\Omega_{\mathrm{b}} h^2$, 
$\Omega_{\mathrm{c}} h^2$, $\theta$, $\tau$, $n_{\mathrm{s}}$, 
$\ln(10^{10} A_{\mathrm{s}})$, $\mathrm{Age}/\mathrm{Gyr}$, $\sigma_8$, 
$z_{\mathrm{re}}$ and $H_0$. Here $\Omega_{\Lambda}$ is the dark energy density today.
$\Omega_{\mathrm{b}}$ is the baryon density today. $\Omega_{\mathrm{c}}$ is the 
cold dark matter density today. $\theta$ is the angular scalar of the sound 
horizon at last-scattering. $\tau$ is the Thomson scattering optical depth due 
to the reionization. $n_{\mathrm{s}}$ is the scalar spectrum power-law index.
$\ln(10^{10} A_{\mathrm{s}})$ is log power of the primordial curvature perturbations.
$\mathrm{Age}/\mathrm{Gyr}$ is the age of the universe. $\sigma_8$ is the rms matter
fluctuations today in linear theory. $z_{\mathrm{re}}$ is the redshift at which
universe is half reionized. $H_0$ is the Hubble constant.

\section{Results}

\begin{figure}[htbp]
\begin{center}
\includegraphics[width=3.2in]{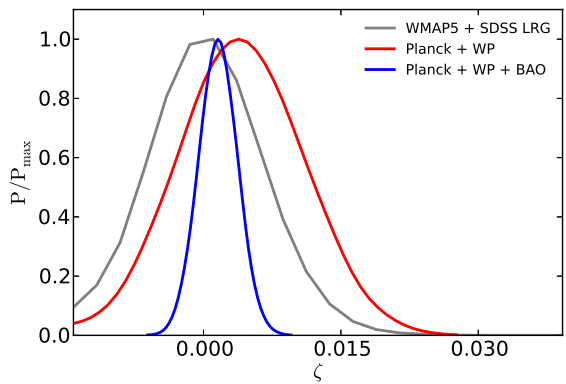}
\caption{The one dimensional likelihood distribution for $\zeta$. 
``Planck + WP" denotes the result of using Planck temperature data as well as
WMAP9 polarization data. ``Planck + WP + BAO" denotes the combined constraint with
BAO data \cite{Percival:2010hx} \cite{Padmanabhan:2012ft} \cite{Anderson:2012uh} 
\cite{Beutler:2011ea}. We also plot the result in previous work \cite{wu:2010brans.constraint}, 
``WMAP5 + SDSS LRG", which is using CMB temperature and polarization data from WMAP5, 
combined with matter power spectrum measured with the luminous red galaxy (LRG) 
survey of the SDSS.}
\label{fig:one_d_likelihood_zeta}
\end{center}
\end{figure}

Fig.\ref{fig:one_d_likelihood_zeta} shows the 1-D marginalized distribution
for the Brans-Dicke parameter $\zeta$. The curve which is labelled as  ``Planck + WP" shows
the CMB-only result, for which the temperature data from Planck and the polarization
data from WMAP9 is used. The curve which is labelled as ``Planck + WP + BAO" combined 
the CMB data with the BAO observation data from SDSS DR7, BOSS DR9 and 6dF. For 
comparison, we also plot in this figure the result obtained in our previous work 
\cite{wu:2010brans.constraint}, which used the WMAP5 data and the matter power spectrum from the 
SDSS LRG survey.  

Comparing with the previous work, especially Wu et al.\cite{wu:2010brans.constraint}, which 
used the same parameterization, the constraints become stronger, and the new data favors
a slightly more positive value of $\zeta$. The likelihood of the CMB-only data looks 
quite Gaussian. Because of the high angular resolution of Planck data, the CMB-only 
data can already give better constraints than before. With the BAO data, 
the constraint is further tightened. 
The BAO distances estimated from galaxy surveys play the same  role as the
matter power spectrum in distinguishing the different models. 

For the CMB-only case, we find 
the 68\% and 95\% intervals are 
\begin{eqnarray}
-0.247\times10^{-2} &< \zeta <& 1.080\times10^{-2} ~~(68\%)\ ; \label{eq:CMBzeta68}\\
-0.855\times10^{-2} &<\zeta <& 1.716\times10^{-2} ~~(95\%)\ .
\end{eqnarray}
These correspond to
\begin{eqnarray}
\omega<-405.36 ~~~~ &\rm{or}& ~~~~ \omega > 92.09 ~~(68\%)\ ; \\
\omega<-117.46 ~~~~ &\rm{or}& ~~~~ \omega > 57.78 ~~(95\%)\ . \label{eq:CMBomega95}
\end{eqnarray}
We see that with the CMB data alone, the constraint is still relatively loose.

Addition of the BAO data helps to break the parameter degeneracy, and much stronger
limits are obtained. For the CMB+BAO case, we find 
the 68\% and 95\% bounds are 
\begin{eqnarray}
-0.046\times10^{-2} < &\zeta& < 0.366\times10^{-2} ~~(68\%)\ ; \label{eq:CMB+BAOzeta68}\\
-0.246\times10^{-2} < &\zeta& < 0.567\times10^{-2} ~~(95\%)\ ,
\end{eqnarray}
which correspond to 
\begin{eqnarray}
\omega<-2174.41 ~~~~ &\rm{or}& ~~~~ \omega > 272.72 ~~(68\%)\ ;\\
\omega<-407.00 ~~~~ &\rm{or}& ~~~~ \omega > 175.87 ~~(95\%)\ . \label{eq:CMB+BAOomega95}
\end{eqnarray}

If we restrict ourselves to the case of $\zeta>0$, or equivalently $\omega >0$, 
then for the CMB only case the 68\% and 95\% bounds are
\begin{eqnarray}
0 &< \zeta <& 0.895 \times 10^{-2} ~~(68\%)\ ; \label{eq:zeta+68} \\ 
0 &< \zeta <& 1.645 \times 10^{-2} ~~(95\%)\ ,
\end{eqnarray}
corresponding to 
\begin{equation}
\omega > 111.23 ~~(68\%); \qquad \omega > 60.29 ~~(95\%)\ . 
\end{equation}
For the CMB+BAO case, 
\begin{eqnarray}
0 &< \zeta <& 0.296 \times 10^{-2} ~~(68\%)\ ; \\
0 &< \zeta <& 0.549 \times 10^{-2} ~~(95\%)\ ,
\end{eqnarray}
corresponding to 
\begin{equation}
\omega > 337.34 ~~(68\%);\qquad \omega > 181.65 ~~(95\%)\ . \label{eq:omega+}
\end{equation}
We see that the constraints are only slightly different 
from their respective positive bounds where 
$\zeta<0 (\omega<0)$ are allowed, even though the {\it a prior} allowed parameter 
space is smaller. This shows that the negative $\zeta$ solutions fit
about also very well, so reduction of parameter space does not
significantly improve the constraint.

As discussed in the last section, these limits depend on the parameterization and 
prior adopted. The results presented here applies to the parameter $\zeta$, 
even though we also quoted limits on $\omega$ since that's what appeared in the 
Brans-Dicke theory. This parameterization is more ``conservative'', so our limits
appeared to be ``weaker'' than Ref.\cite{Avilez:2013tm} even though we have used
the newer and more precise Planck data.

\begin{figure}[htbp]
\begin{center}
\includegraphics[width=3.2in]{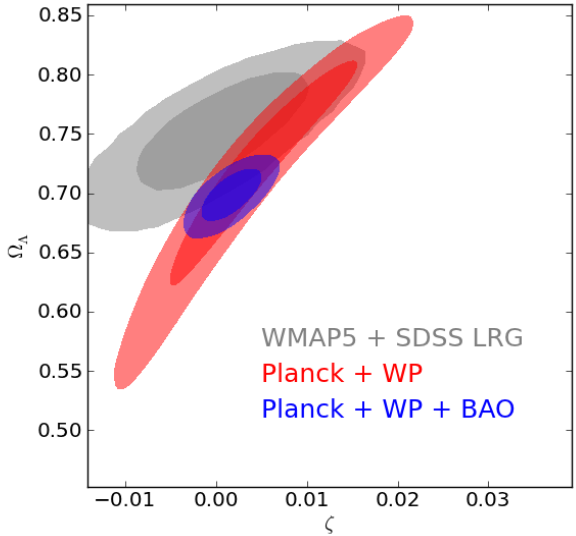}
\caption{The two dimensional contour for $\zeta$ against $\Omega_{\Lambda}$. 
}
\label{fig:two_d_contour_zeta_omegal}
\end{center}
\end{figure}

Fig.\ref{fig:two_d_contour_zeta_omegal} shows the two dimensional contours for $\zeta$ 
against $\Omega_{\Lambda}$. If only the CMB data from Planck is used, the 
constraint already become stronger, but there is significant degeneration between 
$\zeta$ and $\Omega_{\Lambda}$.  The BAO data can help to break the degeneration and 
give much stronger constraints. The center of the contours shifted somewhat from the 
center of our previous results, this is the same trend as seen in the fitting of the 
standard cosmological model, for $\Omega_{\Lambda}$ is lowered, but we see that the 
shift on the center of $\zeta$ is small.

\begin{figure}[htb]
\begin{center}
\includegraphics[width=3.5in]{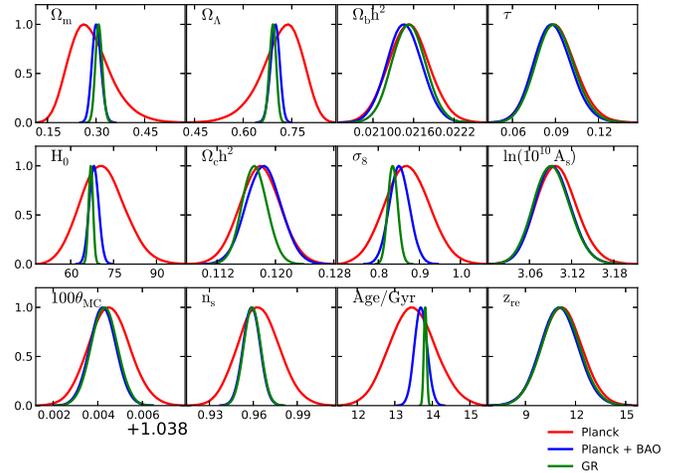}
\caption{The one dimensional likelihood for cosmological parameters. 
The red lines with label ``Planck" represent the result with Planck
temperature data and WMAP9 polarization data; the blue lines with label 
``Planck + BAO" denote the result combined with BAO data; the green line
labelled ``GR''  is the result of fixing $\zeta=0$, in 
which case the Brans-Dicke gravity reduces to Einstein theory.}
\label{fig:one_d_likelihood}
\end{center}
\end{figure}

Next we examine how the constraints on other cosmological parameters are affected if 
we consider the Brans-Dicke gravity. Fig.\ref{fig:one_d_likelihood} shows the one 
dimensional likelihood for some cosmological parameters. In this plot, 
we show the result with $\zeta$ fixed to $0$, labelled as ``GR", 
in this case the Brans-Dicke theory is reduced to the standard $\Lambda CDM$ 
model with Einstein's General Relativity. It is obvious from the figure that when
the BAO data are combined, the constraint is much tighter than the case with CMB data 
only. However, for most parameters, the likelihood distribution of the GR case and 
the Brans-Dicke case is very similar, the shifts in the best fit parameters 
(peak value of the likelihood) are small, and the differences in the width
of the likelihood are also relatively small, showing that the addition of the 
Brans-Dicke parameter does not significantly affect the uncertainty in other cosmological
parameters.  The most affected basic parameters are $H_0$, $\Omega_c h^2$, 
and $\sigma_8$, while for the derived parameters the uncertainty on the cosmic age is 
much larger.

\begin{figure*}[p]
\begin{center}
\includegraphics[width=6in]{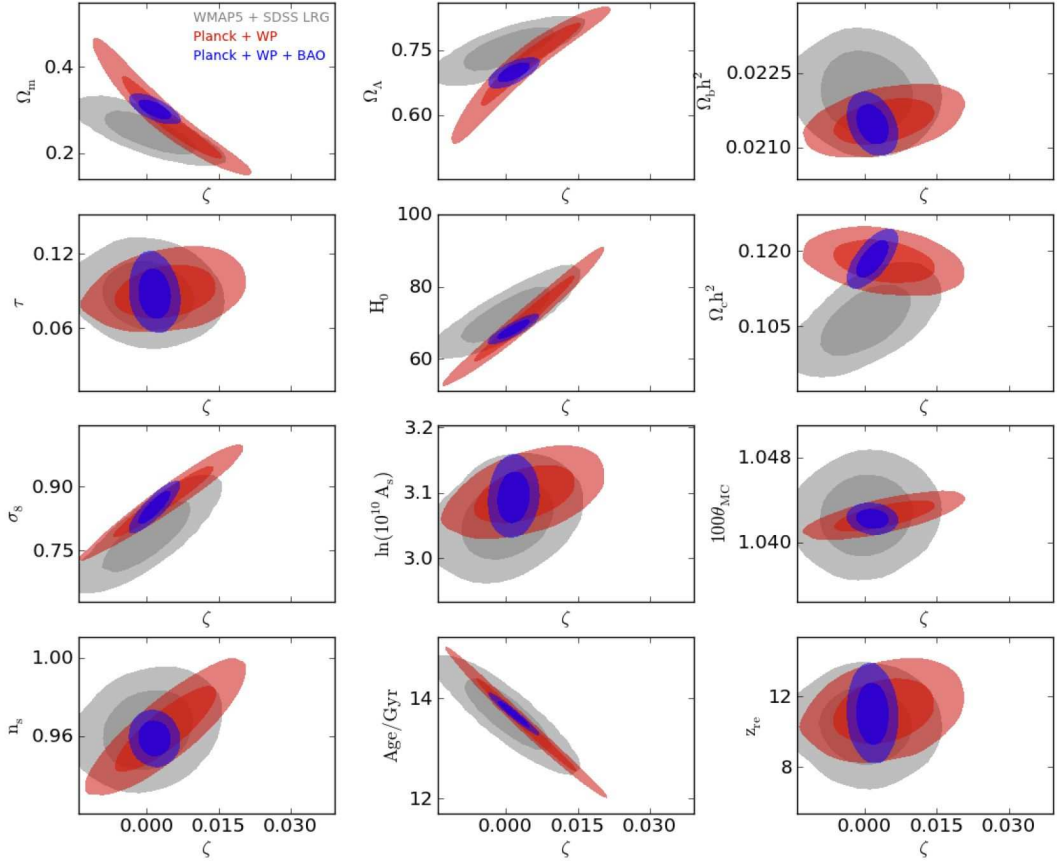}
\caption{The two dimensional contour for cosmological parameters.}
\label{fig:two_d_contour}
\end{center}
\end{figure*}

We also plot the two dimensional
contours in Fig.\ref{fig:two_d_contour}. The Planck data can give accurate 
measure on $\Omega_{\mathrm{b}} h^2$ without other additional data, thanks to the 
well measured peaks in angular power spectra. The degeneracy between $\zeta$ and 
$\Omega_{\mathrm{b}} h^2$, as well as $\Omega_{\mathrm{c}} h^2$ are quite limited 
and the uncertainty of fitting is reduced.  The degeneracy between $\zeta$ 
and $\Omega_m$, $\Omega_\Lambda$ are broken by BAO data, which could also be seen in 
one dimensional likelihood distribution, $\Omega_m$ and $\Omega_\Lambda$ change greatly 
after introducing BAO data.

The best-fit values and $68\%$ marginalized error are shown in Table \ref{tab:params}. 
For comparison,  in Table \ref{tab:params} 
we also list the results for Einstein gravity given in 
Ref.\cite{collaboration:2013vc}. The Einstein result is constrained
by using Planck low-$l$ and high-$l$ data, as well as WMAP9 polarization data and 
BAO data, which is the same as our data set. 
Comparing with the WMAP data, the Planck data favors lower $\Omega_{\Lambda}$ and 
lower $H_0$, in the standard $\Lambda$CDM model fitting. This trend was noted by the 
Planck team and also found in our model. 

There are some slight differences between our results with $\zeta$ fixed to 0 and the 
result published by the Planck team \cite{2013arXiv1303.5076P}. 
The differences come mainly from different setting of parameters in CosmoMC. In our 
fitting, in order to focus on the Brans-Dicke parameters, we ignored the effect of 
massive neutrinos, and fixed the neutrino number. The best-fit values of cosmological 
parameters in our Brans-Dicke model are consistent with $\Lambda CDM$ model in Einstein 
theory.

\begin{figure*}[tb]
\begin{center}
\subfigure[]{
    \includegraphics[width=3.2in]{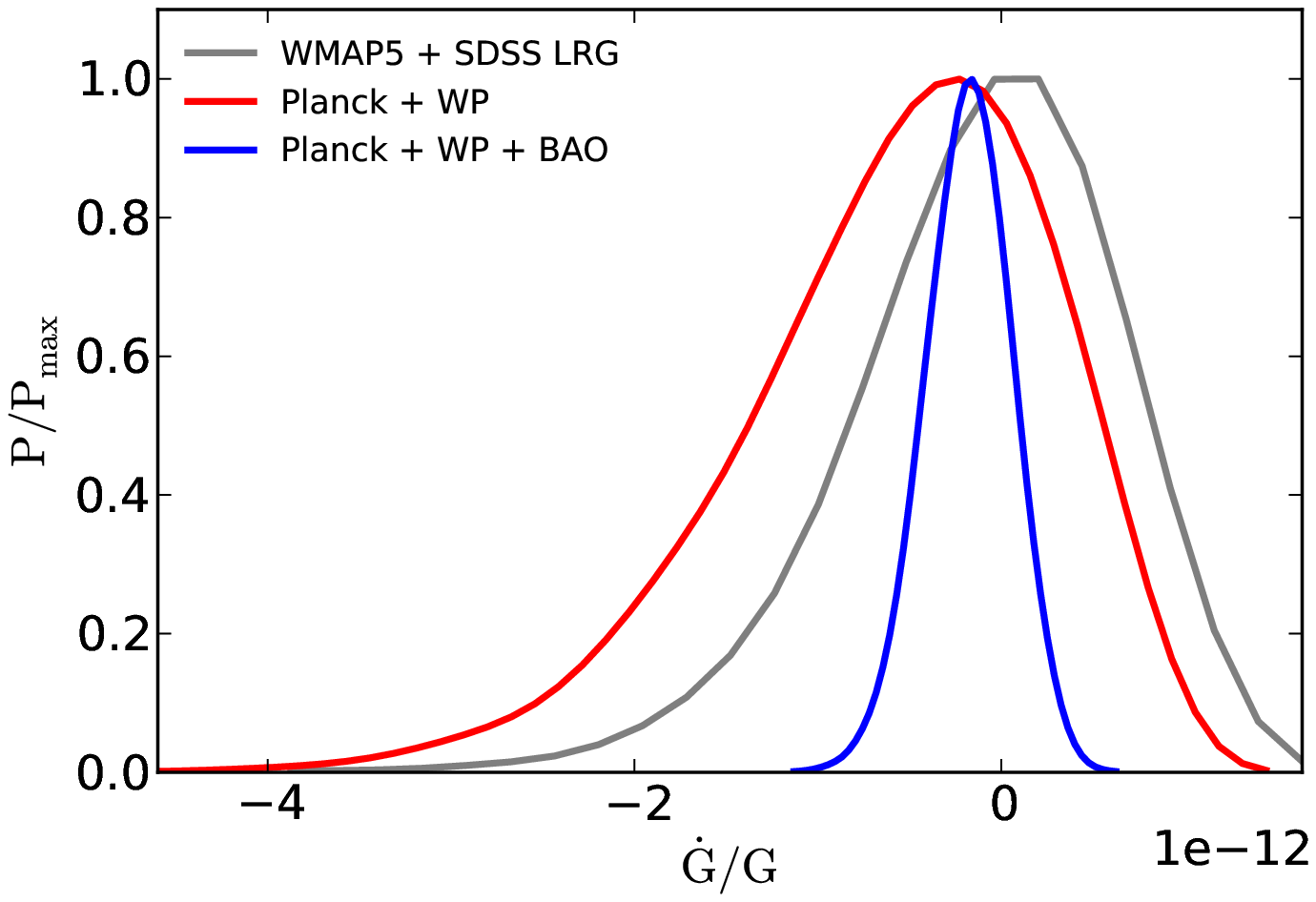}
    \label{fig:one_d_likelihood_gdotog}
    }
\subfigure[]{
    \includegraphics[width=3.2in]{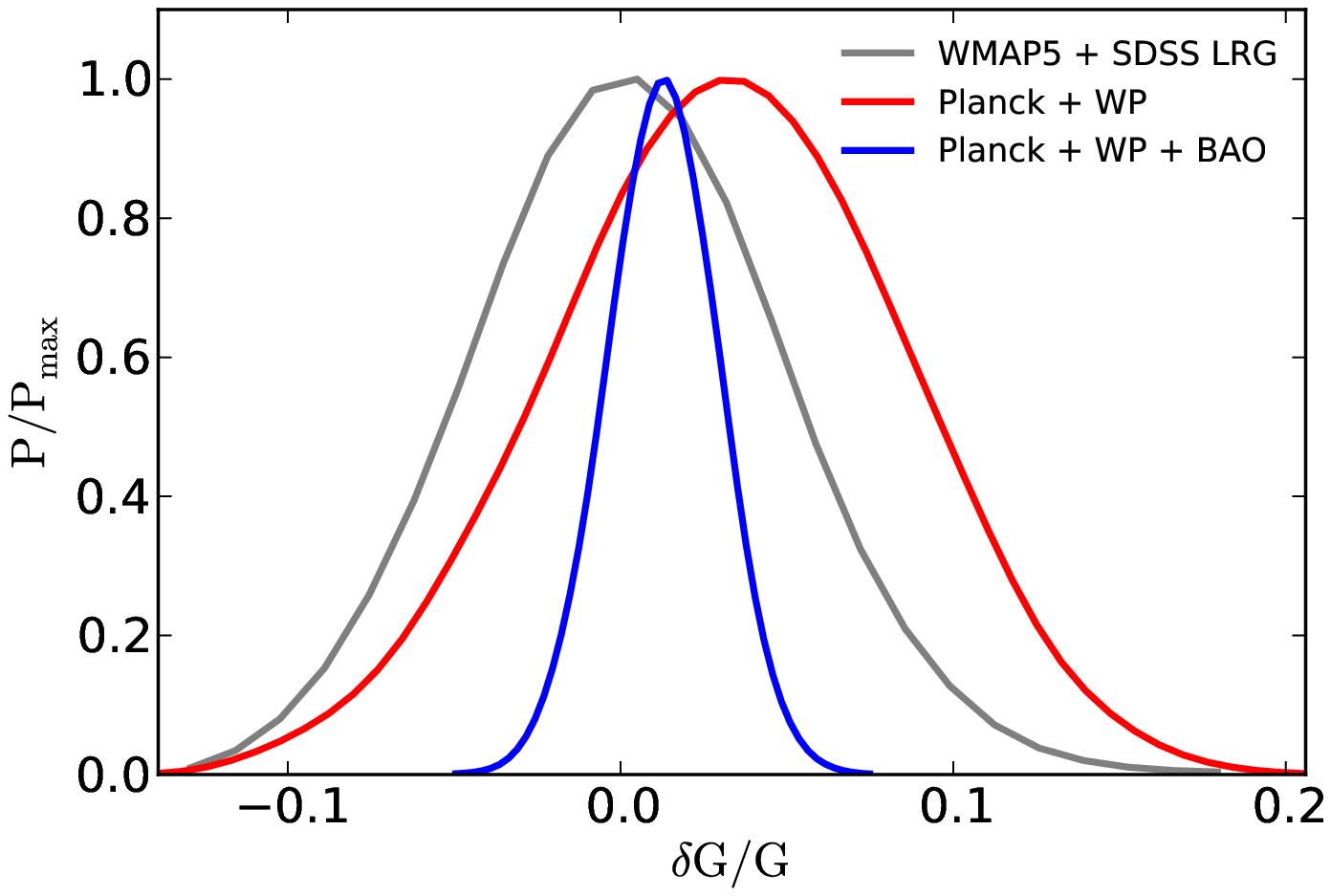}
    \label{fig:one_d_likelihood_dg}
    }    
\caption{Fig.\ref{fig:one_d_likelihood_gdotog}, and Fig.\ref{fig:one_d_likelihood_dg}
show the one dimensional marginalized likelihood of parameters $\dot{G}/G$ and 
$\delta G/G$.
}
\label{fig:gdotog_dg}
\end{center}
\end{figure*}

We can also test the variation of the gravitational constant $G$ in the 
context of Brans-Dicke theory. We added two derived
parameters in the MCMC code, i.e. $\dot{G}/G \equiv -\dot{\varphi}/\varphi$, which is the 
change rate of gravitational constant at present, and 
$\delta G/G \equiv (G_{rec} - G_0)/G_0 $, which is the integrated change of 
gravitational constant since the epoch of recombination. 
The one dimensional marginalized likelihood is shown 
in Fig.\ref{fig:one_d_likelihood_gdotog} and 
Fig.\ref{fig:one_d_likelihood_dg}. The likelihood functions are fairly close to the
Gaussian form. We can take the $68\%$ limit as corresponding to the $1\sigma$ error
for these measurements.

For the CMB only case, the best-fit values are
$$ \dot{G}/G = -0.4617\times10^{-12}, \qquad \delta{G}/G = 0.0318$$ 
and the $68\%$ marginalized limits are
\begin{eqnarray}
-1.1970\times10^{-12} < &\dot{G}/G& < 0.4597\times10^{-12}; \label{eq:GCMB}\\
-0.0197 < &\delta G/G& < 0.0835 
\end{eqnarray}
For the CMB+BAO case, 
$$\dot{G}/G = -0.1417\times10^{-12}, \qquad \delta{G}/G = 0.0104 $$ 
and the $68\%$ marginalized limits are
\begin{eqnarray}
-0.4082\times10^{-12} < &\dot{G}/G& < 0.0663\times10^{-12}  \\
-0.0037 < &\delta G/G& < 0.0290 \label{eq:GCMB+BAO}
\end{eqnarray}
We list the constraints on $\dot{G}/G$ with different methods in Table \ref{tab:G}. 
Though model-dependent, our cosmological constraints are now 
comparable in precision with other methods, including the solar system experiments.

\begin{table}[ht]
\begin{center}
\caption{\label{tab:G} Constraints on the rate of variations of
gravitational constant.  The errors are $1\sigma$ unless otherwise noted. }

{\small
\begin{tabular}{c|c}   \hline\hline
$\dot{G}/G~[10^{-13} $yr$^{-1}]$   & Method \\
 \hline
$2\pm 7$ &lunar laser ranging \cite{Muller:2007zzb}\\
 \hline
$0\pm 4$ &big bang nucleosynthesis \cite{Copi:2003xd}\cite{Bambi:2005fi} \\
 \hline
$0\pm 16$  & helioseismology \cite{Guenther98} \\
 \hline
$-6\pm20$&neutron star mass \cite{Thorsett:1996fr} \\
 \hline
$20\pm40$ &Viking lander ranging \cite{hellings1983} \\
 \hline
$40 \pm 50$ & binary pulsar \cite{Kaspi:1994hp} \\
 \hline
$-96\sim 81 ~(2\sigma)$& CMB (WMAP3) \cite{Chan:2007fe} \\
 \hline
$-17.5\sim 10.5 ~(2\sigma)$& WMAP5+SDSS LRG  \cite{wu:2010brans.constraint} \\
 \hline
$-1.42^{+2.48}_{-2.27} ~(1\sigma) $& Planck+WP+BAO (This paper) \\
\hline
\hline

\end{tabular}
}
\end{center}
\end{table}

\section{Summary}

In this paper, we use the newly published Planck CMB temperature 
data \cite{collaboration:2013vc} and the WMAP 9 year CMB polarization data 
\cite{2012arXiv1212.5225B} to constrain the 
Brans-Dicke theory. In addition to the Planck data, we also use the
BAO data from the SDSS DR7
\cite{Percival:2010hx} \cite{Padmanabhan:2012ft}, BOSS DR9 \cite{Anderson:2012uh} and
6dF\cite{Beutler:2011ea}, which help to break parameter degeneracy.

We use the parameterization $\zeta=\ln(1+\frac{1}{\omega})$ introduced in 
Ref.\cite{wu:2010brans.constraint}, for which the GR limit is achieved when 
 $\zeta \to 0$, ($|\omega| \to \infty$). This parameterization may be 
more ``conservative'' than some of the other parameterizations, so the limit 
we derive may also appear ``weaker'' than given in some of the other works.
The readers should note this when comparing the results given in different works.

We obtained constraints by using the CMB data (referred to as CMB-only). 
The 68\% and 95\% bounds are given 
Eqs.(\ref{eq:CMBzeta68})-(\ref{eq:CMBomega95}).
By combining the BAO data, we obtain stricter 
constraints, which are given in Eqs.(\ref{eq:CMB+BAOzeta68})-(\ref{eq:CMB+BAOomega95}).
We also considered the bounds obtained if $\zeta >0 $ (or equivalently 
$\omega>0$) 
is assumed to be positive, these are given in Eqs.(\ref{eq:zeta+68})-(\ref{eq:omega+}).
We do not detect any significant deviation from Einstein's general theory of relativity,
and the constraint on the Brans-Dicke model is tightened  compared with previous results.

We examined the distribution of other cosmological parameters. For most parameters, 
the best fit values and measurement errors are not altered much by the introduction of the 
Brans-Dicke gravity. The most affected parameters are $H_0$, $\Omega_c h^2$, and 
$\sigma_8$, and the derived parameter of cosmic age.

Finally, the variation of the gravitational constant in the Brans-Dicke model
are also constrained, the results are given in 
Eqs.(\ref{eq:GCMB})-(\ref{eq:GCMB+BAO}), and also summarized in Table \ref{tab:G}.
These constraints are model-dependent, nonetheless, 
it is remarkable that the limits obtained are comparable with the 
constraints from the highly 
precise solar system experiments.

\section*{Acknowledgements}
We thank  Antony Lewis and Xiaoyuan Huang for their helps on 
the CosmoMC code. Our MCMC computation was performed on the Laohu cluster in NAOC.
This work is supported by the Ministry of Science and Technology 863 project 
grant 2012AA121701, the NSFC grant 11073024, 11103027, and the CAS Knowledge 
Innovation grant KJCX2-EW-W01.

\begin{table*}[p]
\begin{center}
\caption{\label{tab:params}Summary of cosmological parameters and the corresponding 
68\% intervals. The ``Planck + WP" column lists the result of using temperature map
from Planck and polarization map from WMAP9; The ``Planck + WP + BAO" column lists
the result with BAO data combined; We also list the result using the same data as
``Planck + WP + BAO", but fix $\zeta=0$ in the ``Planck + WP + BAO with $\zeta=0$"
column, that Brans-Dicke reduces to Einstein theory. The last column is the result
form Planck team in Ref.\cite{collaboration:2013vc}. }
{\scriptsize
\begin{tabular}{c|cc|cc|cc|cc} \hline\hline
& \multicolumn{6}{|c}{Brans-Dicke}& \multicolumn{2}{|c}{Einstein\cite{collaboration:2013vc}}\\
\hline
& \multicolumn{2}{|c}{Planck + WP}& \multicolumn{2}{|c}{Planck + WP + BAO}& \multicolumn{2}{|c}{Planck + WP + BAO with $\zeta=0$}& \multicolumn{2}{|c}{Planck + WP + BAO}\\
\hline
Parameter & Best fit & $68\%$ limits& Best fit & $68\%$ limits& Best fit & $68\%$ limits& Best fit & $68\%$ limits\\
\hline
$\Omega_m$  & $  0.2821$ & $  0.2845_{ -0.0753}^{+ 0.0479}$ & $  0.3048$ & $  0.3016_{ -0.0149}^{+ 0.0133}$ & $  0.3098$ & $  0.3087_{ -0.0110}^{+ 0.0101}$ & &\\
$\Omega_\Lambda$  & $  0.7179$ & $  0.7155_{ -0.0479}^{+ 0.0753}$ & $  0.6952$ & $  0.6984_{ -0.0133}^{+ 0.0149}$ & $  0.6902$ & $  0.6913_{ -0.0101}^{+ 0.0110}$ & $0.6914$ & $ 0.692_{-0.01}^{+0.01}$ \\
$\Omega_{\mathrm{b}} h^2$  & $  0.0215$ & $  0.0215_{ -0.0003}^{+ 0.0003}$ & $  0.0215$ & $  0.0215_{ -0.0003}^{+ 0.0003}$ & $  0.0215$ & $  0.0215_{ -0.0002}^{+ 0.0002}$ & $0.0222$  & $0.0221_{-0.0002}^{+0.0002}$ \\
$\tau$  & $  0.0802$ & $  0.0902_{ -0.0150}^{+ 0.0128}$ & $  0.0871$ & $  0.0883_{ -0.0136}^{+ 0.0122}$ & $  0.0830$ & $  0.0899_{ -0.0137}^{+ 0.0124}$ & $0.0952$ & $0.092_{-0.013}^{+0.013}$ \\
$H_0$  & $ 70.4907$ & $ 71.2328_{ -8.2356}^{+ 7.1229}$ & $ 67.7905$ & $ 68.1442_{ -1.6225}^{+ 1.6147}$ & $ 66.9751$ & $ 67.0443_{ -0.7665}^{+ 0.7621}$ & $67.77$ & $67.80_{-0.77}^{+0.77}$\\
$\Omega_{\mathrm{c}} h^2$  & $  0.1187$ & $  0.1179_{ -0.0027}^{+ 0.0027}$ & $  0.1186$ & $  0.1184_{ -0.0023}^{+ 0.0023}$ & $  0.1174$ & $  0.1172_{ -0.0017}^{+ 0.0017}$ & $0.1189$ & $0.1187_{-0.0017}^{+0.0017}$\\
$\sigma_8$  & $  0.8648$ & $  0.8705_{ -0.0524}^{+ 0.0526}$ & $  0.8507$ & $  0.8519_{ -0.0238}^{+ 0.0239}$ & $  0.8314$ & $  0.8357_{ -0.0125}^{+ 0.0115}$ & $0.8288$ & $0.826_{-0.012}^{+0.012}$\\
$\ln(10^{10} A_\mathrm{s})$  & $  3.0810$ & $  3.0989_{ -0.0305}^{+ 0.0268}$ & $  3.0922$ & $  3.0937_{ -0.0264}^{+ 0.0242}$ & $  3.0797$ & $  3.0921_{ -0.0270}^{+ 0.0245}$ & $3.0973$ & $3.091_{-0.025}^{+0.025}$\\
$100\theta_{\mathrm{MC}}$  & $  1.0423$ & $  1.0425_{ -0.0009}^{+ 0.0009}$ & $  1.0424$ & $  1.0422_{ -0.0006}^{+ 0.0006}$ & $  1.0424$ & $  1.0423_{ -0.0006}^{+ 0.0006}$ & $1.0415$ & $1.0415_{-0.0006}^{+0.0006}$\\
$n_{\mathrm{s}}$  & $  0.9621$ & $  0.9638_{ -0.0137}^{+ 0.0138}$ & $  0.9606$ & $  0.9588_{ -0.0056}^{+ 0.0056}$ & $  0.9584$ & $  0.9593_{ -0.0056}^{+ 0.0056}$ & $0.9611$ & $0.9608_{-0.0054}^{0.0054}$\\
$\mathrm{Age}/\mathrm{Gyr}$  & $ 13.4843$ & $ 13.4730_{ -0.5924}^{+ 0.5892}$ & $ 13.7179$ & $ 13.6921_{ -0.1644}^{+ 0.1637}$ & $ 13.8100$ & $ 13.8119_{ -0.0371}^{+ 0.0371}$ & $13.7965$ & $13.798_{-0.037}^{+0.037}$\\
$z_{\mathrm{re}}$  & $ 10.3531$ & $ 11.1855_{ -1.1589}^{+ 1.1645}$ & $ 10.9543$ & $ 11.0309_{ -1.0862}^{+ 1.0948}$ & $ 10.5331$ & $ 11.1067_{ -1.0984}^{+ 1.0993}$ & $11.52$ & $11.3_{-1.1}^{+1.1}$\\
\hline\hline
\end{tabular}
}
\end{center}
\end{table*}

\bibliography{brans}

\begin{thebibliography}{38}
\expandafter\ifx\csname natexlab\endcsname\relax\def\natexlab#1{#1}\fi
\expandafter\ifx\csname bibnamefont\endcsname\relax
  \def\bibnamefont#1{#1}\fi
\expandafter\ifx\csname bibfnamefont\endcsname\relax
  \def\bibfnamefont#1{#1}\fi
\expandafter\ifx\csname citenamefont\endcsname\relax
  \def\citenamefont#1{#1}\fi
\expandafter\ifx\csname url\endcsname\relax
  \def\url#1{\texttt{#1}}\fi
\expandafter\ifx\csname urlprefix\endcsname\relax\def\urlprefix{URL }\fi
\providecommand{\bibinfo}[2]{#2}
\providecommand{\eprint}[2][]{\url{#2}}

\bibitem[{\citenamefont{Jordan}(1949)}]{Jordan:1949nature}
\bibinfo{author}{\bibfnamefont{P.}~\bibnamefont{Jordan}},
  \bibinfo{journal}{Nature (London)} \textbf{\bibinfo{volume}{164}},
  \bibinfo{pages}{637} (\bibinfo{year}{1949}).

\bibitem[{\citenamefont{Jordan}(1959)}]{Jordan:1959eg}
\bibinfo{author}{\bibfnamefont{P.}~\bibnamefont{Jordan}}, \bibinfo{journal}{Z.
  Phys.} \textbf{\bibinfo{volume}{157}}, \bibinfo{pages}{112}
  (\bibinfo{year}{1959}).

\bibitem[{\citenamefont{Fierz}(1956)}]{Fierz:1956}
\bibinfo{author}{\bibfnamefont{M.}~\bibnamefont{Fierz}},
  \bibinfo{journal}{Helv.\ Phys.\ Acta} \textbf{\bibinfo{volume}{29}},
  \bibinfo{pages}{128} (\bibinfo{year}{1956}).

\bibitem[{\citenamefont{Brans and Dicke}(1961)}]{Brans:1961sx}
\bibinfo{author}{\bibfnamefont{C.}~\bibnamefont{Brans}} \bibnamefont{and}
  \bibinfo{author}{\bibfnamefont{R.~H.} \bibnamefont{Dicke}},
  \bibinfo{journal}{Phys. Rev.} \textbf{\bibinfo{volume}{124}},
  \bibinfo{pages}{925} (\bibinfo{year}{1961}).

\bibitem[{\citenamefont{Dicke}(1962)}]{Dicke:1961gz}
\bibinfo{author}{\bibfnamefont{R.~H.} \bibnamefont{Dicke}},
  \bibinfo{journal}{Phys. Rev.} \textbf{\bibinfo{volume}{125}},
  \bibinfo{pages}{2163} (\bibinfo{year}{1962}).

\bibitem[{\citenamefont{Bergmann}(1968)}]{Bergmann:1968ve}
\bibinfo{author}{\bibfnamefont{P.~G.} \bibnamefont{Bergmann}},
  \bibinfo{journal}{Int. J. Theor. Phys.} \textbf{\bibinfo{volume}{1}},
  \bibinfo{pages}{25} (\bibinfo{year}{1968}).

\bibitem[{\citenamefont{Nordtvedt}(1970)}]{Nordtvedt:1970uv}
\bibinfo{author}{\bibfnamefont{J.}~\bibnamefont{Nordtvedt},
  \bibfnamefont{Kenneth}}, \bibinfo{journal}{Astrophys. J.}
  \textbf{\bibinfo{volume}{161}}, \bibinfo{pages}{1059} (\bibinfo{year}{1970}).

\bibitem[{\citenamefont{Wagoner}(1970)}]{Wagoner:1970vr}
\bibinfo{author}{\bibfnamefont{R.~V.} \bibnamefont{Wagoner}},
  \bibinfo{journal}{Phys. Rev.} \textbf{\bibinfo{volume}{D1}},
  \bibinfo{pages}{3209} (\bibinfo{year}{1970}).

\bibitem[{\citenamefont{Bekenstein}(1977)}]{Bekenstein:1977rb}
\bibinfo{author}{\bibfnamefont{J.~D.} \bibnamefont{Bekenstein}},
  \bibinfo{journal}{Phys. Rev.} \textbf{\bibinfo{volume}{D15}},
  \bibinfo{pages}{1458} (\bibinfo{year}{1977}).

\bibitem[{\citenamefont{Bekenstein and Meisels}(1978)}]{Bekenstein:1978zz}
\bibinfo{author}{\bibfnamefont{J.~D.} \bibnamefont{Bekenstein}}
  \bibnamefont{and} \bibinfo{author}{\bibfnamefont{A.}~\bibnamefont{Meisels}},
  \bibinfo{journal}{Phys. Rev.} \textbf{\bibinfo{volume}{D18}},
  \bibinfo{pages}{4378} (\bibinfo{year}{1978}).

\bibitem[{\citenamefont{{Will}}(1981)}]{Will:1981tegp.book}
\bibinfo{author}{\bibfnamefont{C.~M.} \bibnamefont{{Will}}},
  \emph{\bibinfo{title}{{Theory and experiment in gravitational physics}}}
  (\bibinfo{year}{1981}).

\bibitem[{\citenamefont{Will}(2006)}]{Will:2006wv}
\bibinfo{author}{\bibfnamefont{C.~M.} \bibnamefont{Will}},
  \bibinfo{journal}{Living Reviews in Relativity} \textbf{\bibinfo{volume}{9}},
  \bibinfo{pages}{3} (\bibinfo{year}{2006}).

\bibitem[{\citenamefont{Bertotti et~al.}(2003)\citenamefont{Bertotti, Iess, and
  Tortora}}]{Bertotti:2003by}
\bibinfo{author}{\bibfnamefont{B.}~\bibnamefont{Bertotti}},
  \bibinfo{author}{\bibfnamefont{L.}~\bibnamefont{Iess}}, \bibnamefont{and}
  \bibinfo{author}{\bibfnamefont{P.}~\bibnamefont{Tortora}},
  \bibinfo{journal}{Nature} \textbf{\bibinfo{volume}{425}},
  \bibinfo{pages}{374} (\bibinfo{year}{2003}).

\bibitem[{\citenamefont{Chen and Kamionkowski}(1999)}]{Chen:1999qh}
\bibinfo{author}{\bibfnamefont{X.}~\bibnamefont{Chen}} \bibnamefont{and}
  \bibinfo{author}{\bibfnamefont{M.}~\bibnamefont{Kamionkowski}},
  \bibinfo{journal}{Phys. Rev.} \textbf{\bibinfo{volume}{D60}},
  \bibinfo{pages}{104036} (\bibinfo{year}{1999}), \eprint{astro-ph/9905368}.

\bibitem[{\citenamefont{Acquaviva et~al.}(2005)\citenamefont{Acquaviva,
  Baccigalupi, Leach, Liddle, and Perrotta}}]{Acquaviva:2004ti}
\bibinfo{author}{\bibfnamefont{V.}~\bibnamefont{Acquaviva}},
  \bibinfo{author}{\bibfnamefont{C.}~\bibnamefont{Baccigalupi}},
  \bibinfo{author}{\bibfnamefont{S.~M.} \bibnamefont{Leach}},
  \bibinfo{author}{\bibfnamefont{A.~R.} \bibnamefont{Liddle}},
  \bibnamefont{and} \bibinfo{author}{\bibfnamefont{F.}~\bibnamefont{Perrotta}},
  \bibinfo{journal}{Phys. Rev.} \textbf{\bibinfo{volume}{D71}},
  \bibinfo{pages}{104025} (\bibinfo{year}{2005}), \eprint{astro-ph/0412052}.

\bibitem[{\citenamefont{{Wu} et~al.}(2010)\citenamefont{{Wu}, {Qiang}, {Wang},
  and {Chen}}}]{wu:2010brans.theory}
\bibinfo{author}{\bibfnamefont{F.-Q.} \bibnamefont{{Wu}}},
  \bibinfo{author}{\bibfnamefont{L.-E.} \bibnamefont{{Qiang}}},
  \bibinfo{author}{\bibfnamefont{X.}~\bibnamefont{{Wang}}}, \bibnamefont{and}
  \bibinfo{author}{\bibfnamefont{X.}~\bibnamefont{{Chen}}},
  \bibinfo{journal}{\prd} \textbf{\bibinfo{volume}{82}}, \bibinfo{eid}{083002}
  (\bibinfo{year}{2010}), \eprint{0903.0384}.

\bibitem[{\citenamefont{{Wu} and {Chen}}(2010)}]{wu:2010brans.constraint}
\bibinfo{author}{\bibfnamefont{F.-Q.} \bibnamefont{{Wu}}} \bibnamefont{and}
  \bibinfo{author}{\bibfnamefont{X.}~\bibnamefont{{Chen}}},
  \bibinfo{journal}{\prd} \textbf{\bibinfo{volume}{82}}, \bibinfo{eid}{083003}
  (\bibinfo{year}{2010}), \eprint{0903.0385}.

\bibitem[{\citenamefont{Avilez and Skordis}(2013)}]{Avilez:2013tm}
\bibinfo{author}{\bibfnamefont{A.}~\bibnamefont{Avilez}} \bibnamefont{and}
  \bibinfo{author}{\bibfnamefont{C.}~\bibnamefont{Skordis}}
  (\bibinfo{year}{2013}).

\bibitem[{\citenamefont{{Bennett} et~al.}(2012)\citenamefont{{Bennett},
  {Larson}, {Weiland}, {Jarosik}, {Hinshaw}, {Odegard}, {Smith}, {Hill},
  {Gold}, {Halpern} et~al.}}]{2012arXiv1212.5225B}
\bibinfo{author}{\bibfnamefont{C.~L.} \bibnamefont{{Bennett}}},
  \bibinfo{author}{\bibfnamefont{D.}~\bibnamefont{{Larson}}},
  \bibinfo{author}{\bibfnamefont{J.~L.} \bibnamefont{{Weiland}}},
  \bibinfo{author}{\bibfnamefont{N.}~\bibnamefont{{Jarosik}}},
  \bibinfo{author}{\bibfnamefont{G.}~\bibnamefont{{Hinshaw}}},
  \bibinfo{author}{\bibfnamefont{N.}~\bibnamefont{{Odegard}}},
  \bibinfo{author}{\bibfnamefont{K.~M.} \bibnamefont{{Smith}}},
  \bibinfo{author}{\bibfnamefont{R.~S.} \bibnamefont{{Hill}}},
  \bibinfo{author}{\bibfnamefont{B.}~\bibnamefont{{Gold}}},
  \bibinfo{author}{\bibfnamefont{M.}~\bibnamefont{{Halpern}}},
  \bibnamefont{et~al.}, \bibinfo{journal}{ArXiv e-prints}
  (\bibinfo{year}{2012}), \eprint{1212.5225}.

\bibitem[{\citenamefont{{Hinshaw} et~al.}(2012)\citenamefont{{Hinshaw},
  {Larson}, {Komatsu}, {Spergel}, {Bennett}, {Dunkley}, {Nolta}, {Halpern},
  {Hill}, {Odegard} et~al.}}]{2012arXiv1212.5226H}
\bibinfo{author}{\bibfnamefont{G.}~\bibnamefont{{Hinshaw}}},
  \bibinfo{author}{\bibfnamefont{D.}~\bibnamefont{{Larson}}},
  \bibinfo{author}{\bibfnamefont{E.}~\bibnamefont{{Komatsu}}},
  \bibinfo{author}{\bibfnamefont{D.~N.} \bibnamefont{{Spergel}}},
  \bibinfo{author}{\bibfnamefont{C.~L.} \bibnamefont{{Bennett}}},
  \bibinfo{author}{\bibfnamefont{J.}~\bibnamefont{{Dunkley}}},
  \bibinfo{author}{\bibfnamefont{M.~R.} \bibnamefont{{Nolta}}},
  \bibinfo{author}{\bibfnamefont{M.}~\bibnamefont{{Halpern}}},
  \bibinfo{author}{\bibfnamefont{R.~S.} \bibnamefont{{Hill}}},
  \bibinfo{author}{\bibfnamefont{N.}~\bibnamefont{{Odegard}}},
  \bibnamefont{et~al.}, \bibinfo{journal}{ArXiv e-prints}
  (\bibinfo{year}{2012}), \eprint{1212.5226}.

\bibitem[{\citenamefont{{Planck Collaboration}
  et~al.}(2013{\natexlab{a}})\citenamefont{{Planck Collaboration}, {Ade},
  {Aghanim}, {Armitage-Caplan}, {Arnaud}, {Ashdown}, {Atrio-Barandela},
  {Aumont}, {Baccigalupi}, {Banday} et~al.}}]{2013arXiv1303.5062P}
\bibinfo{author}{\bibnamefont{{Planck Collaboration}}},
  \bibinfo{author}{\bibfnamefont{P.~A.~R.} \bibnamefont{{Ade}}},
  \bibinfo{author}{\bibfnamefont{N.}~\bibnamefont{{Aghanim}}},
  \bibinfo{author}{\bibfnamefont{C.}~\bibnamefont{{Armitage-Caplan}}},
  \bibinfo{author}{\bibfnamefont{M.}~\bibnamefont{{Arnaud}}},
  \bibinfo{author}{\bibfnamefont{M.}~\bibnamefont{{Ashdown}}},
  \bibinfo{author}{\bibfnamefont{F.}~\bibnamefont{{Atrio-Barandela}}},
  \bibinfo{author}{\bibfnamefont{J.}~\bibnamefont{{Aumont}}},
  \bibinfo{author}{\bibfnamefont{C.}~\bibnamefont{{Baccigalupi}}},
  \bibinfo{author}{\bibfnamefont{A.~J.} \bibnamefont{{Banday}}},
  \bibnamefont{et~al.}, \bibinfo{journal}{ArXiv e-prints}
  (\bibinfo{year}{2013}{\natexlab{a}}), \eprint{1303.5062}.

\bibitem[{\citenamefont{collaboration et~al.}(2013)\citenamefont{collaboration,
  Ade, Aghanim, Armitage-Caplan, Arnaud, Ashdown, Atrio-Barandela, Aumont,
  Baccigalupi, Banday et~al.}}]{collaboration:2013vc}
\bibinfo{author}{\bibfnamefont{P.}~\bibnamefont{collaboration}},
  \bibinfo{author}{\bibfnamefont{P.~A.~R.} \bibnamefont{Ade}},
  \bibinfo{author}{\bibfnamefont{N.}~\bibnamefont{Aghanim}},
  \bibinfo{author}{\bibfnamefont{C.}~\bibnamefont{Armitage-Caplan}},
  \bibinfo{author}{\bibfnamefont{M.}~\bibnamefont{Arnaud}},
  \bibinfo{author}{\bibfnamefont{M.}~\bibnamefont{Ashdown}},
  \bibinfo{author}{\bibfnamefont{F.}~\bibnamefont{Atrio-Barandela}},
  \bibinfo{author}{\bibfnamefont{J.}~\bibnamefont{Aumont}},
  \bibinfo{author}{\bibfnamefont{C.}~\bibnamefont{Baccigalupi}},
  \bibinfo{author}{\bibfnamefont{A.~J.} \bibnamefont{Banday}},
  \bibnamefont{et~al.} (\bibinfo{year}{2013}).

\bibitem[{\citenamefont{{Planck Collaboration}
  et~al.}(2013{\natexlab{b}})\citenamefont{{Planck Collaboration}, {Ade},
  {Aghanim}, {Armitage-Caplan}, {Arnaud}, {Ashdown}, {Atrio-Barandela},
  {Aumont}, {Baccigalupi}, {Banday} et~al.}}]{2013arXiv1303.5076P}
\bibinfo{author}{\bibnamefont{{Planck Collaboration}}},
  \bibinfo{author}{\bibfnamefont{P.~A.~R.} \bibnamefont{{Ade}}},
  \bibinfo{author}{\bibfnamefont{N.}~\bibnamefont{{Aghanim}}},
  \bibinfo{author}{\bibfnamefont{C.}~\bibnamefont{{Armitage-Caplan}}},
  \bibinfo{author}{\bibfnamefont{M.}~\bibnamefont{{Arnaud}}},
  \bibinfo{author}{\bibfnamefont{M.}~\bibnamefont{{Ashdown}}},
  \bibinfo{author}{\bibfnamefont{F.}~\bibnamefont{{Atrio-Barandela}}},
  \bibinfo{author}{\bibfnamefont{J.}~\bibnamefont{{Aumont}}},
  \bibinfo{author}{\bibfnamefont{C.}~\bibnamefont{{Baccigalupi}}},
  \bibinfo{author}{\bibfnamefont{A.~J.} \bibnamefont{{Banday}}},
  \bibnamefont{et~al.}, \bibinfo{journal}{ArXiv e-prints}
  (\bibinfo{year}{2013}{\natexlab{b}}), \eprint{1303.5076}.

\bibitem[{\citenamefont{Percival et~al.}(2010)\citenamefont{Percival, Reid,
  Eisenstein, Bahcall, Budav{\'a}ri, Frieman, Fukugita, Gunn, Ivezi{\'c}, Knapp
  et~al.}}]{Percival:2010hx}
\bibinfo{author}{\bibfnamefont{W.~J.} \bibnamefont{Percival}},
  \bibinfo{author}{\bibfnamefont{B.~A.} \bibnamefont{Reid}},
  \bibinfo{author}{\bibfnamefont{D.~J.} \bibnamefont{Eisenstein}},
  \bibinfo{author}{\bibfnamefont{N.~A.} \bibnamefont{Bahcall}},
  \bibinfo{author}{\bibfnamefont{T.}~\bibnamefont{Budav{\'a}ri}},
  \bibinfo{author}{\bibfnamefont{J.~A.} \bibnamefont{Frieman}},
  \bibinfo{author}{\bibfnamefont{M.}~\bibnamefont{Fukugita}},
  \bibinfo{author}{\bibfnamefont{J.~E.} \bibnamefont{Gunn}},
  \bibinfo{author}{\bibfnamefont{{\v Z}.}~\bibnamefont{Ivezi{\'c}}},
  \bibinfo{author}{\bibfnamefont{G.~R.} \bibnamefont{Knapp}},
  \bibnamefont{et~al.}, \bibinfo{journal}{Monthly Notices of the Royal
  Astronomical Society} \textbf{\bibinfo{volume}{401}}, \bibinfo{pages}{2148}
  (\bibinfo{year}{2010}).

\bibitem[{\citenamefont{Padmanabhan et~al.}(2012)\citenamefont{Padmanabhan, Xu,
  Eisenstein, Scalzo, Cuesta, Mehta, and Kazin}}]{Padmanabhan:2012ft}
\bibinfo{author}{\bibfnamefont{N.}~\bibnamefont{Padmanabhan}},
  \bibinfo{author}{\bibfnamefont{X.}~\bibnamefont{Xu}},
  \bibinfo{author}{\bibfnamefont{D.~J.} \bibnamefont{Eisenstein}},
  \bibinfo{author}{\bibfnamefont{R.}~\bibnamefont{Scalzo}},
  \bibinfo{author}{\bibfnamefont{A.~J.} \bibnamefont{Cuesta}},
  \bibinfo{author}{\bibfnamefont{K.~T.} \bibnamefont{Mehta}}, \bibnamefont{and}
  \bibinfo{author}{\bibfnamefont{E.}~\bibnamefont{Kazin}},
  \bibinfo{journal}{Monthly Notices of the Royal Astronomical Society}
  \textbf{\bibinfo{volume}{427}}, \bibinfo{pages}{2132} (\bibinfo{year}{2012}).

\bibitem[{\citenamefont{Anderson et~al.}(2012)\citenamefont{Anderson, Aubourg,
  Bailey, Bizyaev, Blanton, Bolton, Brinkmann, Brownstein, Burden, Cuesta
  et~al.}}]{Anderson:2012uh}
\bibinfo{author}{\bibfnamefont{L.}~\bibnamefont{Anderson}},
  \bibinfo{author}{\bibfnamefont{E.}~\bibnamefont{Aubourg}},
  \bibinfo{author}{\bibfnamefont{S.}~\bibnamefont{Bailey}},
  \bibinfo{author}{\bibfnamefont{D.}~\bibnamefont{Bizyaev}},
  \bibinfo{author}{\bibfnamefont{M.}~\bibnamefont{Blanton}},
  \bibinfo{author}{\bibfnamefont{A.~S.} \bibnamefont{Bolton}},
  \bibinfo{author}{\bibfnamefont{J.}~\bibnamefont{Brinkmann}},
  \bibinfo{author}{\bibfnamefont{J.~R.} \bibnamefont{Brownstein}},
  \bibinfo{author}{\bibfnamefont{A.}~\bibnamefont{Burden}},
  \bibinfo{author}{\bibfnamefont{A.~J.} \bibnamefont{Cuesta}},
  \bibnamefont{et~al.} (\bibinfo{year}{2012}).

\bibitem[{\citenamefont{Beutler et~al.}(2011)\citenamefont{Beutler, Blake,
  Colless, Jones, Staveley-Smith, Campbell, Parker, Saunders, and
  Watson}}]{Beutler:2011ea}
\bibinfo{author}{\bibfnamefont{F.}~\bibnamefont{Beutler}},
  \bibinfo{author}{\bibfnamefont{C.}~\bibnamefont{Blake}},
  \bibinfo{author}{\bibfnamefont{M.}~\bibnamefont{Colless}},
  \bibinfo{author}{\bibfnamefont{D.~H.} \bibnamefont{Jones}},
  \bibinfo{author}{\bibfnamefont{L.}~\bibnamefont{Staveley-Smith}},
  \bibinfo{author}{\bibfnamefont{L.}~\bibnamefont{Campbell}},
  \bibinfo{author}{\bibfnamefont{Q.}~\bibnamefont{Parker}},
  \bibinfo{author}{\bibfnamefont{W.}~\bibnamefont{Saunders}}, \bibnamefont{and}
  \bibinfo{author}{\bibfnamefont{F.}~\bibnamefont{Watson}},
  \bibinfo{journal}{Monthly Notices of the Royal Astronomical Society}
  \textbf{\bibinfo{volume}{416}}, \bibinfo{pages}{3017} (\bibinfo{year}{2011}).

\bibitem[{\citenamefont{Lewis and Challinor}(1999)}]{CAMB}
\bibinfo{author}{\bibfnamefont{A.}~\bibnamefont{Lewis}} \bibnamefont{and}
  \bibinfo{author}{\bibfnamefont{A.}~\bibnamefont{Challinor}},
  \emph{\bibinfo{title}{http://camb.info/}} (\bibinfo{year}{1999}).

\bibitem[{\citenamefont{Lewis and Bridle}(2002)}]{Lewis:2002ah}
\bibinfo{author}{\bibfnamefont{A.}~\bibnamefont{Lewis}} \bibnamefont{and}
  \bibinfo{author}{\bibfnamefont{S.}~\bibnamefont{Bridle}},
  \bibinfo{journal}{Phys. Rev.} \textbf{\bibinfo{volume}{D66}},
  \bibinfo{pages}{103511} (\bibinfo{year}{2002}), \eprint{astro-ph/0205436}.

\bibitem[{\citenamefont{{Caldwell} et~al.}(2003)\citenamefont{{Caldwell},
  {Kamionkowski}, and {Weinberg}}}]{2003PhRvL..91g1301C}
\bibinfo{author}{\bibfnamefont{R.~R.} \bibnamefont{{Caldwell}}},
  \bibinfo{author}{\bibfnamefont{M.}~\bibnamefont{{Kamionkowski}}},
  \bibnamefont{and} \bibinfo{author}{\bibfnamefont{N.~N.}
  \bibnamefont{{Weinberg}}}, \bibinfo{journal}{Physical Review Letters}
  \textbf{\bibinfo{volume}{91}}, \bibinfo{eid}{071301} (\bibinfo{year}{2003}),
  \eprint{arXiv:astro-ph/0302506}.

\bibitem[{\citenamefont{Muller and Biskupek}(2007)}]{Muller:2007zzb}
\bibinfo{author}{\bibfnamefont{J.}~\bibnamefont{Muller}} \bibnamefont{and}
  \bibinfo{author}{\bibfnamefont{L.}~\bibnamefont{Biskupek}},
  \bibinfo{journal}{Class. Quant. Grav.} \textbf{\bibinfo{volume}{24}},
  \bibinfo{pages}{4533} (\bibinfo{year}{2007}).

\bibitem[{\citenamefont{Copi et~al.}(2004)\citenamefont{Copi, Davis, and
  Krauss}}]{Copi:2003xd}
\bibinfo{author}{\bibfnamefont{C.~J.} \bibnamefont{Copi}},
  \bibinfo{author}{\bibfnamefont{A.~N.} \bibnamefont{Davis}}, \bibnamefont{and}
  \bibinfo{author}{\bibfnamefont{L.~M.} \bibnamefont{Krauss}},
  \bibinfo{journal}{Phys. Rev. Lett.} \textbf{\bibinfo{volume}{92}},
  \bibinfo{pages}{171301} (\bibinfo{year}{2004}), \eprint{astro-ph/0311334}.

\bibitem[{\citenamefont{Bambi et~al.}(2005)\citenamefont{Bambi, Giannotti, and
  Villante}}]{Bambi:2005fi}
\bibinfo{author}{\bibfnamefont{C.}~\bibnamefont{Bambi}},
  \bibinfo{author}{\bibfnamefont{M.}~\bibnamefont{Giannotti}},
  \bibnamefont{and} \bibinfo{author}{\bibfnamefont{F.~L.}
  \bibnamefont{Villante}}, \bibinfo{journal}{Phys. Rev.}
  \textbf{\bibinfo{volume}{D71}}, \bibinfo{pages}{123524}
  (\bibinfo{year}{2005}), \eprint{astro-ph/0503502}.

\bibitem[{\citenamefont{{Guenther} et~al.}(1998)\citenamefont{{Guenther},
  {Krauss}, and {Demarque}}}]{Guenther98}
\bibinfo{author}{\bibfnamefont{D.~B.} \bibnamefont{{Guenther}}},
  \bibinfo{author}{\bibfnamefont{L.~M.} \bibnamefont{{Krauss}}},
  \bibnamefont{and}
  \bibinfo{author}{\bibfnamefont{P.}~\bibnamefont{{Demarque}}},
  \bibinfo{journal}{Astrophys. J.} \textbf{\bibinfo{volume}{498}},
  \bibinfo{pages}{871} (\bibinfo{year}{1998}).

\bibitem[{\citenamefont{Thorsett}(1996)}]{Thorsett:1996fr}
\bibinfo{author}{\bibfnamefont{S.~E.} \bibnamefont{Thorsett}},
  \bibinfo{journal}{Phys. Rev. Lett.} \textbf{\bibinfo{volume}{77}},
  \bibinfo{pages}{1432} (\bibinfo{year}{1996}), \eprint{astro-ph/9607003}.

\bibitem[{\citenamefont{{Hellings} et~al.}(1983)\citenamefont{{Hellings},
  {Adams}, {Anderson}, {Keesey}, {Lau}, {Standish}, {Canuto}, and
  {Goldman}}}]{hellings1983}
\bibinfo{author}{\bibfnamefont{R.~W.} \bibnamefont{{Hellings}}},
  \bibinfo{author}{\bibfnamefont{P.~J.} \bibnamefont{{Adams}}},
  \bibinfo{author}{\bibfnamefont{J.~D.} \bibnamefont{{Anderson}}},
  \bibinfo{author}{\bibfnamefont{M.~S.} \bibnamefont{{Keesey}}},
  \bibinfo{author}{\bibfnamefont{E.~L.} \bibnamefont{{Lau}}},
  \bibinfo{author}{\bibfnamefont{E.~M.} \bibnamefont{{Standish}}},
  \bibinfo{author}{\bibfnamefont{V.~M.} \bibnamefont{{Canuto}}},
  \bibnamefont{and}
  \bibinfo{author}{\bibfnamefont{I.}~\bibnamefont{{Goldman}}},
  \bibinfo{journal}{Physical Review Letters} \textbf{\bibinfo{volume}{51}},
  \bibinfo{pages}{1609} (\bibinfo{year}{1983}).

\bibitem[{\citenamefont{Kaspi et~al.}(1994)\citenamefont{Kaspi, Taylor, and
  Ryba}}]{Kaspi:1994hp}
\bibinfo{author}{\bibfnamefont{V.~M.} \bibnamefont{Kaspi}},
  \bibinfo{author}{\bibfnamefont{J.~H.} \bibnamefont{Taylor}},
  \bibnamefont{and} \bibinfo{author}{\bibfnamefont{M.~F.} \bibnamefont{Ryba}},
  \bibinfo{journal}{Astrophys. J.} \textbf{\bibinfo{volume}{428}},
  \bibinfo{pages}{713} (\bibinfo{year}{1994}).

\bibitem[{\citenamefont{Chang and Chu}(2007)}]{Chan:2007fe}
\bibinfo{author}{\bibfnamefont{K.-C.} \bibnamefont{Chang}} \bibnamefont{and}
  \bibinfo{author}{\bibfnamefont{M.~C.} \bibnamefont{Chu}},
  \bibinfo{journal}{Phys. Rev.} \textbf{\bibinfo{volume}{D75}},
  \bibinfo{pages}{083521} (\bibinfo{year}{2007}), \eprint{astro-ph/0611851}.

\end{thebibliography}
\bibliographystyle{apsrev}
\end{document}